\documentstyle{article}
\begin{document}
\begin{center}
{\bf The Capacity of Quantum Channel with General Signal States}
\vskip30pt
{\bf A.S.Holevo}\\

{\small Steklov Mathematical Institute,\\ Vavilova 42, 117966 Moscow, Russia\\

(e-mail: HOLEVO@CLASS.MI.RAS.RU)}
\end{center}
\begin{abstract} {\small It is shown that the capacity of a classical-quantum
channel  with arbitrary (possibly mixed) states equals to the maximum of the
entropy bound with respect to all apriori distributions.  This completes the
recent result of Hausladen, Jozsa, Schumacher, Westmoreland and Wooters
\cite{jozsa}, who proved the equality for the pure state channel.}
\end{abstract}
\vskip15pt
\noindent
{\bf 1. Information and capacity for quantum channel}.  We start by repeating
some definitions and results from  \cite{hol79}.  Let $\cal H$ be a
$d$-dimensional Hilbert space. We denote $D = \{1,...,d\}$.  A simple quantum
communication channel ({\sl classical-quantum channel} in terminology of
\cite{hol77}) consists of the input alphabet $A = \{1,...,a\}$ and a mapping
$i\rightarrow S_i$ from the input alphabet to the set of quantum states in
$\cal H$. A quantum state is a density operator (d. o.), i. e.  positive
operator $S$ in $\cal H$ with unit trace, Tr$S = 1$. {\sl Coding} is a
probability distribution $\pi = \{\pi_i\}$ on $A$.  {\sl Decoding} is a
resolution of identity in $\cal H$, i. e. a family $X = \{ X_j\}$ of positive
operators in $\cal H$ satisfying $\sum_j X_j = I$, where $I$ is the unit
operator in $\cal H$. The index $j$ runs through some finite output alphabet,
which is not fixed here.  The conditional probability of the output $j$ if the
input was $i$ equals to $P(j|i) = \mbox{Tr}S_i X_j$. The Shannon information is
given by the classical formula $$I_1 (\pi , X ) = \sum_j \sum_i \pi_i
P(j|i)\mbox{log}\left( \frac{P(j|i)} {\sum_k \pi_k P(j|k)}\right), \eqno(1)$$
(in what follows we use the binary logarithms).
 
In the same way we can consider the product channel in ${\cal H}^{\otimes n} =
{\cal H}\otimes ...\otimes {\cal H}$ with the input alphabet $A^n$ consisting
of words $u = (i_1 ,...,i_n )$ of length $n$, with the d. o.  $$S_u =
S_{i_1}\otimes ...\otimes S_{i_n}  \eqno (2)$$ corresponding to the word $u$.
If ${\pi}$ is a probability distribution on $A^n$ and $X$ is a resolution of
identity in ${\cal H}^{\otimes n}$, we define the information quantity $I_n
(\pi , X )$ by the formula similar to (1).  Defining $$C_n = \sup_{\pi , X} I_n
( \pi ,X),$$ we have the property of superadditivity $C_n + C_m \leq C_{n+m}$,
hence the following limit exists $$C = \lim_{n \to \infty}C_n /n , \eqno(3)$$
and is called the {\sl capacity} of the initial channel \cite{hol79}. This
definition is justified by the fact easily deduced from the classical Shannon's
coding theorem, that $C$ is the least upper bound of rate (bits/symbol) of
information which can be transmitted with asymptotically vanishing error.  More
precisely, we call by {\sl code of size} $N$ a sequence $(u_1 , X_1 ),..., (u_n
, X_n )$, where $u_k$ are words of length $n$, and $\{ X_k \}$ is a family of
positive operators in ${\cal H}^{\otimes n}$, satisfying $\sum_{j=1}^N X_j \leq
I$. Defining $X_0 = I - \sum_{j=1}^N X_j$, we have a resolution of identity in
${\cal H}^{\otimes n}$.  An output $k (1\leq k\leq N)$ means decision that the
word $u_k$ was transmitted, while the output $0$ is interpreted as evasion of
any decision.The average error probability for such a code is $${\sf P}_{er} =
\frac{1}{N} \sum_{k=1}^N [1 - \mbox{Tr}S_{u_k} X_k ]. $$ Let us denote $p(n,
N)$ the minimum of this error probability with respect to all codes of the size
$N$ with words of length $n$.  Then  $$p(n, 2^{n(C - \delta )}) \rightarrow
0\qquad \mbox{and} \qquad p(n, 2^{n(C + \delta )}) \not\rightarrow 0,
\eqno(4)$$ where $\delta > 0,$ if $  n \rightarrow \infty$.  The same holds for
the minimum of the maximal (with respect to $k$) error probability, which does
not presume any apriori probabilities for the words (see \cite{inf}, 
\cite{hol79}).
 
{\bf 2. The entropy bound.} The main result of \cite{hol79} was a lower bound
for $C$ demonstrating the
possibility of the inequality $C > C_1$ and implying  strict superadditivity of
the sequence $C_n$.  This is in sharp contrast with the situation for the
corresponding classical memoriless channel, for which $C_n = n C_1$ and hence
$C = C_1$, and is just another manifestation of the quantum nonseparability.
This fact is in a sense dual to the existence of EPR correlations: the latter
are due to entangled states and hold for disentangled measurements while the
superadditivity is due to entangled measurements and holds for
disentangled states. The inequality $C\not= C_1$ raised the problem of the
actual value of the capacity $C$.

Let $H(S) = - \mbox{Tr}S\mbox{log}S$ be the von Neumann entropy of a d. o.  $S$
and let $\pi = \{\pi_i \}$ be an apriori distribution on $A$. Let us denote
${\bar S} = \sum_{i\in A}\pi_i S_i , {\bar H}(S_{(\cdot )}) =
\sum_{i\in A}\pi_i H(S_i )$ and $\Delta H (\pi ) = H({\bar S}) - 
 {\bar H}(S_{(\cdot )})$. The entropy bound \cite{hol73} combined with an
additivity  property  proved in Appendix implies $C \leq \max_{\pi}\Delta H(\pi
).$ In \cite{hol79} a conjecture was made that in fact this might be an
equality. Recently Hausladen, Jozsa, Schumacher, Westmoreland and Wooters
\cite{jozsa} proved this in the case of pure states $S_i$
(apparently not knowing about the paper \cite{hol79}). The problem for the case
of general (possibly mixed) states was left open and is the subject of our
present work.  The main result is the estimate for the error probability
implying converse inequality $C \geq \max_{\pi}\Delta H(\pi )$. Thus we have

{\bf Theorem}.~ {\sl The capacity of the quantum communication channel with
arbitrary signal states $S_i$ is given by} $$C = \max_{\pi} [H(\sum_{i\in
A}\pi_i S_i ) - \sum_{i\in A}\pi_i H(S_i )], \eqno(5)$$ confirming the old
physical wisdom according to which the entropy bound was used to evaluate the
quantum capacity \cite{gordon}.

The key points of the proof are the idea of projection onto the typical 
subspace due to
\cite{schum}, \cite{jozsa}, modified here for the case of mixed states,  and the
estimate for the error probability, which is substantially more complicated
than the estimate for pure states given already in \cite{hol79} and a similar
estimate from
\cite{jozsa}.

{\bf 3. The typical subspaces of density operators.} Let ${\bar S} = \sum_{j\in
D}\lambda_j |e_j >\\<e_j |$ be the spectral decomposition of the d. o. ${\bar
S}$, then the spectral decomposition of ${\bar S}^{\otimes n} = {\bar S}
\otimes ...\otimes {\bar S}$ is $${\bar
S}^{\otimes n} = \sum_{J\in D^n}\lambda_J |e_J ><e_J |,$$ where $J = (j_1
,...,j_n ),\quad \lambda_J = \lambda_{j_1} \cdot ... \cdot
\lambda_{j_n},\quad |e_J > = |e_{j_1}>\otimes ...\otimes |e_{j_n}>.$ Following
\cite{jozsa} we introduce the spectral projector onto the {\sl typical subspace}
of the d. o. ${\bar S}^{\otimes n}$ as $$P= \sum_{J\in B} |e_J ><e_J |,   \eqno
(6)$$ where $B = \{J: 2^{-n[H({\bar S})+\delta ]} < \lambda_J < 2^{-n[H({\bar
S})-\delta ]}\}
\subset D^n$ . A sequence $J\in B$ is
``typical'' for a probability distribution on $D^n$ given by eigenvalues
$\lambda_J$ of the d. o. ${\bar S}^{\otimes n}$ in the sense of classical
information theory (see e. g. \cite{inf}). It follows that for fixed small
positive $\epsilon ,\delta$ and all $n\geq n_1 (\pi,\epsilon ,\delta )$
$$\mbox{Tr}{\bar S}^{\otimes n}(I - P)\leq\epsilon .   \eqno(7)$$ Indeed,
$\mbox{Tr}{\bar S}^{\otimes n}P$ is equal to the probability $${\sf P}\{ J\in
B\} = {\sf P}\{ n[H({\bar S}) - \delta ] < - \mbox{log}\lambda_J < n[H({\bar
S}) + \delta ]\}$$ $$ = {\sf P} \{ |n^{-1}\sum_{l=1}^{n}\mbox{log}\lambda_{j_l}
+ H({\bar S})| < \delta \},$$ which tends to 1 as $ n\rightarrow \infty $,
according to the Law of Large Numbers, since $H({\bar S}) = - {\sf
M}\mbox{log}\lambda_{(\cdot )}$.

The next step is a developement of this idea necessary to prove the Theorem for
mixed states. Let $S_i = \sum_{j\in D}\lambda_j^i |e_j^i ><e_j^i |$ be the
spectral decomposition of the d. o. $S_i$. Let $u = (i_1 ,..., i_n )$ be a word
of the input alphabet and $S_u = S_{i_1}\otimes ...\otimes S_{i_n}$ be the
corresponding d. o.  Its spectral decomposition is $$S_u = \sum_{J\in
D^n}\lambda_J^u |e_J^u ><e_J^u |,$$ where $ \lambda_J^u = \lambda_{j_1}^{i_1}
\cdot ... \cdot
\lambda_{j_n}^{i_n}, |e_J^u > = |e_{j_1}^{i_1}>\otimes ...\otimes 
|e_{j_n}^{i_n}>.$ We introduce the spectral projector onto the typical
subspace of $S_u$ as $$P_u = \sum_{J\in B_u } |e_J^u ><e_J^u |,   \eqno (8)$$
where $B_u = \{J: 2^{-n[{\bar H}(S_{(\cdot )})+\delta ]} < \lambda_J^u <
2^{-n[{\bar H}(S_{(\cdot )})-\delta ]}\}.$

Let on the set of all words $A^n$ the following probability distribution be
defined $${\sf P}\{u=(i_1 ,...,i_n )\} = \pi_{i_1}\cdot ...\cdot\pi_{i_n }.
\eqno(9)$$ Then for fixed small positive $\epsilon ,\delta$ and all $n\geq n_2
(\pi,\epsilon ,\delta )$ $${\sf M}\mbox{Tr}S_u (I - P_u )\leq\epsilon.
\eqno(10)$$ Indeed, consider the sequence of independent trials with the
outcomes $i_l ,j_l ; l = 1,...,n$ where the probability of the outcome $(i, j)$
in each trial is equal to $\pi_i \lambda_j^i$. Then $${\sf M}\mbox{Tr}S_u P_u =
{\sf P}\{ J\in B_u \} = {\sf P}\{ n[{\bar H}(S_{(\cdot )}) -\delta ] < -
\mbox{log}\lambda_J^u < n[{\bar H}(S_{(\cdot )})+\delta ]\}$$ $$ = {\sf P} \{
|n^{-1}\sum_{l=1}^{n}
\mbox{log}\lambda_{j_l}^{i_l}
+ {\bar H}(S_{(\cdot )})| < \delta \},$$ which tends to 1 as $n\rightarrow
\infty $, according to the Law of Large Numbers, since ${\bar H}(S_{(\cdot )})
= - {\sf M}\mbox{log}\lambda_{(\cdot )}^{(\cdot )}$.  In what follows we put $n
(\pi,\epsilon ,\delta ) = \max\{ n_1 (\pi,\epsilon ,\delta ),\\ n_2 (\pi,\epsilon
,\delta )\}.$

{\bf 4. The choice of the suboptimal decision rule.} Let $u_1 ,..., u_N $ be a
sequence of words.  To simplify notations we denote the words by their numbers
$1,...,N$. Put $$X_u = (\sum_{u'=1}^N PP_{u'}P)^{-\frac{1}{2}}PP_u P
(\sum_{u'=1}^N PP_{u'}P)^{-\frac{1}{2}},  \eqno(11)$$ where $X^{-\frac{1}{2}}$
denotes generalized inverse of the operator $X^{\frac{1}{2}}$ i. e. operator
equal $0$ on Ker$X$ and $(X^{\frac{1}{2}})^{-1}$ on Ker$X^{\bot}$. Then
$\sum_{u=1}^N X_u \leq I$. Put $|{\hat e}_J^u > = P|e_J^u >$ where $P$ is
defined by (6), then $$X_u = (\sum_{u'=1}^N \sum_{J\in B_{u'} } |{\hat
e}_J^{u'} ><{\hat e}_J^{u'} |)^{-\frac{1}{2}}
\sum_{J\in B_u } |{\hat e}_J^u ><{\hat e}_J^u |
(\sum_{u'=1}^N \sum_{J\in B_{u'} } |{\hat e}_J^{u'} ><{\hat e}_J^{u'}
|)^{-\frac{1}{2}}.$$ By denoting $$\alpha_{(u,J),(u',J')} = <{\hat e}_J^u|
(\sum_{u'=1}^N \sum_{J\in B_{u''} } |{\hat e}_J^{u''} ><{\hat e}_J^{u''}
|)^{-\frac{1}{2}}\quad {\hat e}_{J'}^{u'}>,$$ and taking into account that
$X_u = P X_u P$, the average error probability
corresponding to the choice (11) can be written as $${\sf P}_{er} =
\frac{1}{N}\sum_{u=1}^N [1 - \sum_{J\in D^n}\sum_{J'\in B_u}
\lambda_J^u |\alpha_{(u,J),(u,J')}|^2 ].    \eqno(12)$$

{\bf 5. The estimate for the error probability.} Taking into account that
$\sum_{J\in D^n }\lambda_J^u = 1$ and omitting some nonpositive terms, we see
that $${\sf P}_{er} \leq \frac{1}{N}\sum_{u=1}^N [\sum_{J\in B_u}
\lambda_J^u (1 - \alpha_{(u,J),(u,J)} ^2) + \sum_{J\not\in B_u }
\lambda_J^u ].    \eqno(13)$$
Let us denote $$\gamma_{(u,J),(u',J')} = <{\hat e}_J^u| {\hat
e}_{J'}^{u'}> = < e_J^u| P e_{J'}^{u'}>  \eqno(14)$$ and introduce the Gram
matrix $$\Gamma = [ \gamma_{(u,J),(u',J')} ],$$ where $J\in B_u,
J'\in B_{u'}$ and $u, u' = 1,...,N$. Then $$\Gamma^{\frac{1}{2}} = [
\alpha_{(u,J),(u',J')} ].$$ In particular, $\alpha_{(u,J),(u,J)} ^2 \leq
\gamma_{(u,J),(uJ)} \leq 1$.  Then from (13) $${\sf P}_{er} \leq
\frac{1}{N}\sum_{u=1}^N [2\sum_{J\in B_u}
\lambda_J^u (1 - \alpha_{(u,J),(u,J)}) + \sum_{J\not\in B_u }
\lambda_J^u ].    \eqno(15)$$

By introducing the diagonal matrix $\Lambda = \mbox{diag}[\lambda_J^u ]$ and
denoting by $E$ the unit matrix and the trace of matrices by Sp as distinct
from the trace of operators in Hilbert space, we have $$2\sum_{u=1}^N 
\sum_{J\in B_u}\lambda_J^u (1 - \alpha_{(u,J),(u,J)}) = 2\mbox{Sp}\Lambda 
(E - \Gamma^{\frac{1}{2}})$$
$$= \mbox{Sp}\Lambda (E - \Gamma^{\frac{1}{2}})^2  + \mbox{Sp}\Lambda (E -
\Gamma )
\leq \mbox{Sp}\Lambda (E - \Gamma )^2 + \mbox{Sp}\Lambda (E - \Gamma ) \eqno(16)$$
since $(E - \Gamma^{\frac{1}{2}})^2 = (E - \Gamma )^2 (E +
\Gamma^{\frac{1}{2}})^{-2} \leq (E - \Gamma )^2$ \cite{hol79}.  Calculating the
traces, we obtain the right hand side of (16) as $$\sum_{u=1}^N \sum_{J\in
B_u}\lambda_J^u [2 - 3\gamma_{(u,J),(u,J)} + \gamma_{(u,J),(u,J)}^2$$ 
$$ + \sum_{J': J'\not= J}|\gamma_{(u,J),(u,J')}|^2 + \sum_{u': u'\not= u}\sum_{J'\in B_{u'}}
|\gamma_{(u,J),(u',J')}|^2 ].$$ This quantity will not decrease if the range of
$J$ is enlarged to the full range $D^n$ and if $2 - 3\gamma_{(u,J),(u,J)}
+\gamma_{(u,J),(u,J)}^2$ is replaced with $2 - 2\gamma_{(u,J),(u,J)}$. Then we
obtain $${\sf P}_{er} \leq \frac{1}{N}\sum_{u=1}^N \{ \sum_{J\in D^n}
\lambda_J^u [2 - 2 \gamma_{(u,J),(u,J)} + 
\sum_{J': J'\not= J}|\gamma_{(u,J),(u,J')}|^2 $$ $$+ \sum_{u': u'\not= u}
\sum_{J'\in B_{u'}}
|\gamma_{(u,J),(u',J')}|^2 ] + \sum_{J\not\in B_u }\lambda_J^u \}.$$

Taking into account the definition (14) of $\gamma_{(u,J),(u',J')}$ and the
fact that $< e_J^u|  e_{J'}^{u}> = 0$ for $J \not= J'$, we can write the
last inequality as $${\sf P}_{er} \leq \frac{1}{N}\sum_{u=1}^N \{ 2 \mbox{Tr}S_u
(I - P) + \mbox{Tr}S_u (I - P)P_u (I - P)$$ $$
+ \sum_{u': u'\not= u} \mbox{Tr}P S_u P P_{u'}  + \mbox{Tr}S_u (I - P_u)\}.$$
The second term is less or equal than $\mbox{Tr}S_u (I - P)$. Thus, finally
$${\sf P}_{er} \leq \frac{1}{N}\sum_{u=1}^N \{ 3\mbox{Tr}S_u (I - P) +
\sum_{u': u'\not= u} \mbox{Tr}P S_u P P_{u'}  
+ \mbox{Tr}S_u (I - P_u)\}. \eqno(17)$$

{\bf 6. The random coding.} Let us assume that the words $u_1 ,..., u_N$ are
chosen at random, independently and with the probability distribution (9) for
each word. Then ${\sf M}S_u = {\bar S}^{\otimes n}$ \cite{jozsa} and from (17),
by independence of $S_u ,P_{u'}$, 
$${\sf M}{\sf P}_{er} \leq  3\mbox{Tr} {\bar S}^{\otimes n} (I - P) + 
(N-1)\mbox{Tr}P {\bar S}^{\otimes n} P {\sf M}P_{u'} + {\sf M}\mbox{Tr}S_u 
(I - P_u).$$ By the
inequalities (7), (10) and by the properties of trace, $${\sf M}{\sf P}_{er}
\leq  4\epsilon + (N-1) \| P {\bar S}^{\otimes n} P\|\mbox{Tr} {\sf M}P_{u'},$$
for $n\geq n (\pi,\epsilon ,\delta )$.  By the definition of $P$, $$\|P {\bar
S}^{\otimes n} P\| \leq 2^{-n[H({\bar S}) - \delta ]},$$ and by the definition
of $P_u$, $$\mbox{Tr} {\sf M}P_{u'} =  {\sf M}\mbox{Tr} P_{u'} \leq {\sf
M}\mbox{Tr} S_{u'} \cdot 2^{n[{\bar H}(S_{(\cdot )}) + \delta ]} = 2^{n[{\bar
H}(S_{(\cdot )}) + \delta ]}.$$ Thus $${\sf M}{\sf P}_{er} \leq  4\epsilon +
(N-1) 2^{-n[H({\bar S}) - {\bar H}(S_{(\cdot )}) - 2\delta ]}. \eqno(18)$$

Let us choose the distribution $\pi = \pi^0$ maximizing the entropy bound
$\Delta H(\pi )$. Then (18) implies $$p(n, N)\leq  4\epsilon + (N-1)
2^{-n[\Delta H(\pi^0 ) - 2\delta ]} \eqno(19)$$ for $n\geq n(\pi^0 , \epsilon
,\delta )$.  Thus $p(n, 2^{n[\Delta H(\pi^0 ) - 3\delta ]})\rightarrow 0$ as $n
\rightarrow
\infty$, whence $\Delta H(\pi^0 ) - 3\delta \leq C$ by (4) for arbitrary
$\delta$, and (5) follows.
\vskip10pt
{\small {\sl Acknowledgements}. The work was stimulated by discussions with
Profs. R. Jozsa and A. Yu. Kitaev during the 3d Conference on Quantum
Communication and Measurement in Hakone, Japan, September 1996, 
where the result of
\cite{jozsa} was reported.  The author is grateful to Prof. O. Hirota, Drs. M.
Osaki and M. Sasaki (Tomogawa University) for their hospitality and stimulating
discussions. Financial support from Tamagawa University, JSPS and RFBR grant N
96-01-01709 is acknowledged.}
\vskip10pt
{\bf Appendix.} Let $A_k , k=1,2,$ be finite alphabets and let $\{S_i^k , i\in
A^k\}$ be families of d. o. in Hilbert spaces ${\cal H}_k .$ Let $\{\pi_{ij}\}$
be a probability distribution on $A^1 \times A^2$ and denote $\Delta
H(\{\pi_{ij}\} ) = H(\sum_{ij}\pi_{ij}S_i^1 \otimes S_j^2 ) -
\sum_{ij}\pi_{ij}H(S_i^1 \otimes S_j^2 )$.  We wish to prove $$\max_{\pi_{ij}}
\Delta H(\{\pi_{ij}\} ) = \max_{\pi_i^1 }
\Delta H(\{\pi_{i}^1 \} )+ \max_{\pi_i^2 }\Delta H(\{\pi_{i}^2 \} ).$$
 By the property of entropy $$H(S)\leq H(\mbox{Tr}_2 S) + H(\mbox{Tr}_1 S),$$
where $S$ is a d. o. in ${\cal H}_1 \otimes{\cal H}_2$ and  $\mbox{Tr}_k S,
k=1,2,$ is partial trace with respect to ${\cal H}_k$, proved in \cite{araki},
we have $$H(\sum_{ij}\pi_{ij}S_i^1 \otimes S_j^2 )\leq H(\sum_{i}\pi_{i}^1
S_i^1 ) + H(\sum_{j}\pi_{j}^2 S_j^2 ),$$ where $\{\pi_i^1\}, \{\pi_i^2\}$ are
the marginal distributions of $\{\pi_{ij}
\}$. It follows that
$$\max_{\pi_{ij}} \Delta H(\{\pi_{ij}\} )\leq \max_{\pi_i^1 }
\Delta H(\{\pi_{i}^1 \} )+ \max_{\pi_i^2 }\Delta H(\{\pi_{i}^2 \} ).$$
The converse inequality follows by restricting to $\pi_{ij}=\pi_{i}^1 \times
\pi_{i}^2$ and using the additivity of quantum entropy for product states.

\end{document}